\documentstyle[prl,aps,epsfig]{revtex}
\begin{document}

\draft
\flushbottom
\twocolumn[\hsize\textwidth\columnwidth\hsize
\csname@twocolumnfalse\endcsname

\title{
Insulator-Metal Phase Diagram of the Optimally Doped Manganites\\
from the Disordered Holstein-Double Exchange Model
}

\author{Sanjeev Kumar$^*$ and Pinaki Majumdar }

\address{ Harish-Chandra  Research Institute,\\
 Chhatnag Road, Jhusi, Allahabad 211 019, India }

\date{Oct 24,  2005}

\maketitle
\tightenlines
\widetext
\advance\leftskip by 57pt
\advance\rightskip by 57pt
\begin{abstract}

We study the Holstein-Double Exchange  model in  three dimensions in the
presence of substitutional disorder. Using a new Monte Carlo technique we 
establish the phase diagram of the clean model  and then focus on the effect 
of varying electron-phonon  coupling and disorder at fixed electron 
density.  We demonstrate how extrinsic 
disorder  controls the interplay of lattice polaron effects and  spin 
fluctuations and leads to widely varying regimes in transport.  Our  results 
on the disorder dependence of the ferromagnetic $T_c$ and metal-insulator 
transitions bear direct comparison to data on the 
`optimally doped', $x=0.3-0.4$, manganites. We  highlight disorder induced 
polaron formation as a key effect in these materials,   
organise a wide variety of data into a simple `global phase diagram',
and make several experimental predictions.

\

\

\end{abstract}

]

\narrowtext
\tightenlines

The Holstein-Double Exchange (H-DE) model provides a minimal description 
of strongly coupled charge, spin, and lattice degrees of freedom, 
typical of the perovskite 
manganites A$_{1-x}$A$'_x$MnO$_3$ 
\cite{mang-genl-ref}, and captures
the interplay of polaronic tendency and spin disorder that 
occurs in these materials.  Although a detailed solution has not been
available in three dimensions (3D), the H-DE model is believed to describe 
the trend towards stronger localisation  and weakened ferromagnetism
observed with reducing A site ionic radius $(r_A)$ in the manganites.

The role of quenched disorder in these systems, 
arising, for example, from the cation size
mismatch $(\sigma_A)$ is only partially  understood. 
Most of the disorder related 
theory \cite{dag-disord,disord-furu} has focused 
on effects near a first order 
phase boundary, {\it i.e}, the `bicritical' region \cite{tok-bicr}, 
motivated by the observation of cluster coexistence 
in some materials \cite{cluster-cmr-expt}.
Phase competetion and bicriticality, 
however, occurs only in a limited 
part of the phase diagram. Studies 
with controlled variation in $\sigma_A$  
indicate that quenched disorder has a dramatic effect  
on the ferromagnetic  $T_c$  
and resistivity \cite{attf-lide,attf-sum,liu,coey}, 
and optical response  \cite{saitoh} 
even far from bicriticality.  
The origin of these effects remain~unexplained.

Our primary contribution in this paper is
to demonstrate how this unusual sensitivity 
arises due to a positive feedback of quenched disorder 
on the polaron formation tendency, and its interplay with
spin fluctuations. 
Through a detailed real space solution of the disordered H-DE model
on large 3D lattices we $(i)$~demonstrate disorder controlled  
metal-insulator transitions (MIT),
in the ground state and  at finite temperature,  
that are absent in the `clean' system, 
$(ii)$~explain the rapid suppression of  $T_c$ 
with increasing disorder, 
$(iii)$~compare our results in detail with data on the manganites
at $x=0.3-0.4$,  constructing a  `global phase diagram', and 
$(iv)$~make testable  predictions about the $\sigma_A$ 
dependence of optical spectral weight, lattice distortion, and
tunneling density of states in the optimally doped manganites.

We study the  adiabatic disordered $(d)$ H-DE model: 
\begin{eqnarray}
H & =& -t\sum_{\langle ij \rangle \sigma}  
(~c^{\dagger}_{i \sigma} c^{~}_{j \sigma} ~+~h.c~)
+  \sum_{i }(\epsilon_i - \mu) n_i  \cr
&& ~~~~~- \lambda \sum_i n_i  x_i  
-J_H\sum_i {\bf S}_i.{\sigma}_i + {K \over 2}\sum_i x_i^2
\end{eqnarray}
The $t$ are nearest neighbour hopping on a cubic lattice, 
$\epsilon_i$
is the quenched  
binary  disorder, with ${\bar \epsilon_i} =0$ and
value $\pm \Delta$, $\lambda$ is the
electron-phonon (EP)
interaction, coupling electron density to the 
local distortion $x_i$, $K$ is the stiffness, 
$\mu$ is the chemical potential, and
$J_H$ is the Hunds coupling.
The parameters in the problem are $\lambda/t$, $\Delta/t$,   
electron density $n$, and temperature $T$.
We treat the phonons and spins as classical (the adiabatic limit), 
assume $J_H/t \rightarrow \infty$,
and set
$\vert {\bf S} \vert =1$. 
We also set $K=1$ and   measure energy, frequency, $T$, {\it  etc}, in
units of $t$.

The clean H-DE model was proposed early on \cite{roder-prl}
as a minimal model for the manganites, and a  
mean field study \cite{roder-prl} suggested strong decrease
of $T_c$ and increasing localisation with increase in EP coupling. 
More recently the H-DE model with 
`cooperative phonons' has been analysed \cite{hde-mc}
in 3D, yielding a thermally
driven MIT at low $n$ but not near optimality. 
The effect of disorder  has been considered on EP-DE models
but with focus mainly on phase coexistence \cite{dag-2band}
and bicriticality \cite{disord-furu}. These studies yield insight
on how disorder leads to cluster formation near a first order
phase boundary \cite{dag-disord,dag-2band}, or its impact on a commensurate
charge ordered phase \cite{disord-furu}, but do not have much
bearing on the effects observed, for example, at $x=0.3$, where there
is no obvious phase competetion.
In our understanding, the effects of
disorder in the manganites, away from bicriticality,
await an explanation.
In this paper we overcome a key technical limitation of the earlier studies
and provide new  results on transport combining the effects 
of strong EP coupling, thermal spin fluctuations, and quenched disorder, 
on lattices upto $12^3$ in size.

Since we work with $J_H/t \rightarrow \infty$, the electron spin is `slaved'
to the core spin orientation leading to an effectively spinless, hopping
disordered model:
$ H  = -\sum_{\langle ij \rangle }  
(t_{ij}\gamma^{\dagger}_i \gamma_j ~+~h.c~)
+  \sum_{i }(\epsilon_i - \mu -\lambda x_i) n_i  + (K/2) \sum_i x_i^2.$
The hopping amplitude depends on the spin orientations via
$t_{ij}/t = cos{\theta_i \over 2}cos{\theta_j \over 2}$~ 
$+$ 
$~sin{\theta_i \over 2}sin{\theta_j \over 2}e^{i(\phi_i - \phi_j)}$, 
where $\theta_i$
and $\phi_i$ are respectively the polar and azimuthal angles of the spin
${\bf S}_i$.
We use our recently developed  ``travelling cluster approximation''
(TCA) \cite{tca-ref}
 for solving the annealing problem on large lattices, by computing 
the energy cost of a MC update by constructing a `cluster Hamiltonian', 
of $N_c = L_c^3$ sites, around the 
update site. 
We will use $N_c = 4^3$ and, as documented earlier
\cite{tca-ref}, this
is sufficient for equilibriating the spin and lattice variables.
The electronic properties, {\it e.g}, 
resistivity, density of states,
and optical response, are obtained by diagonalising the {\it full} electron
Hamiltonian on the 
equilibrium $\{ x_i, {\bf S}_i\}$ configurations,
averaging thermally and 
finally over $8-10$ 
 realisations of disorder $\{ \epsilon_i \}$. 
The  transport results are obtained using a 
formulation described in detail previously \cite{sk-pm-scr}.
We will use system size $N = 8^3$, but have checked our results on
$N$ upto $12^3$.
Our resistivity results are in units of ${\hbar a_0}/(\pi e^2)$, 
where $a_0$ is the lattice spacing. As a rough estimate, $\rho=1$ in our units
would correspond to $\sim 30 \mu \Omega$cm.

Let us summarise the phases of the clean H-DE model before 
addressing 
the effects of disorder.
$({\bf a}).$~Structurally,
 a phase can be  either charge ordered (CO), or
charge disordered, {\it i.e} a polaron liquid (PL), or
homogeneous, {\it i.e} a Fermi liquid (FL). 
$({\bf b}).$~Magnetically it can
be either ferromagnetic (F)
or paramagnetic (P), and $({\bf c}).$~in 
terms of electrical response it can be a `metal',  
with $d\rho/dT >0$, 
or an insulator, with  $d\rho/dT <0$.  

Combining these features, the ground 
state can be either 
a ferromagnetic Fermi liquid 
(F-FL), or F-PL, or F-CO, 
depending on $\lambda$ and $n$,
while finite $T$ allows 
P-FL, P-PL, and P-CO phases as well.
Notice that while  the 
threshold 
for {\it single polaron} trapping, {\it i.e} $n=0$, in 3D, 
is 
$E_p = \lambda^2/2K \approx 5.4t$, 
{\it i.e}  $\lambda_c(n=0) \sim 3.3$  \cite{sk-pm-latt-pol2},
at {\it finite 
density}
the threshold for a collective PL is lower, 
{\it e.g}, $\lambda_c(n=0.3) \sim 2.6$. 
These thresholds are further affected
\begin{center}
\begin{figure}
\epsfxsize=7.0cm \epsfysize=3.7cm \epsfbox{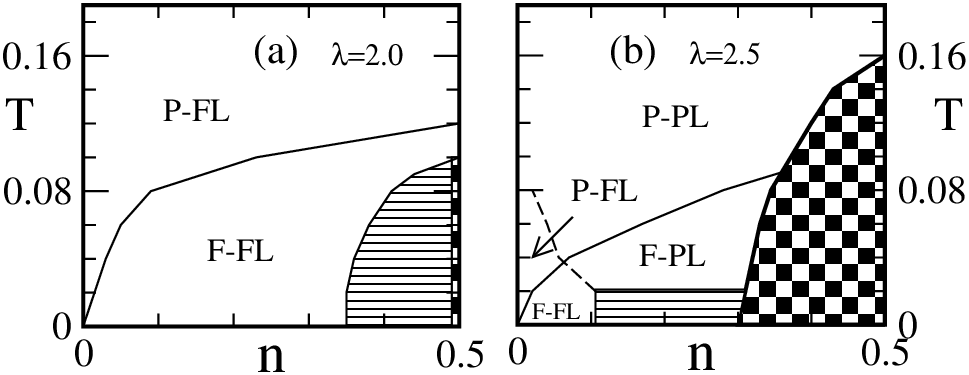}
\vspace{.5mm}
\caption{Phase diagram  in the clean problem:
$(a)$ and $(b)$ show the phase diagram at $\lambda=2.0$ and
$\lambda=2.5$ respectively. The notation for the phases is explained in the
text. The chessboard pattern is for a charge ordered (CO) phase with
$(\pi, \pi, \pi)$ modulation. Shaded regions represent~coexistence.}
\end{figure}
\end{center}
by structural and spin disorder.

For  $\lambda \lesssim  2.0$, Fig.1.$(a)$,
the phases are  $(i)$~F-FL  at low $T$  away from $n=0.5$,
$(ii)$~P-FL  at high $T$, and  $(iii)$~a F-CO  phase  at $n=0.5$, 
separated
from the F-FL,  at low $T$,  by a window of phase separation. 
For  $\lambda=2.0$ the
$T_c$ at $n=0.5$ is suppressed with respect to
the pure DE problem, but is still  
larger than the CO  scale
$T_{CO}$.  At stronger coupling, of which $\lambda=2.5$ is
typical, Fig.1.$(b).$, 
much of the phase diagram is taken up by insulating phases.
As  $\lambda$ is increased beyond $2.5$ the metallic window at
low $n$ shrinks quickly and all the phases turn insulating. The
ground state, however,  continues to be ferromagnetic since we
have not included any competing antiferromagnetism (AF).
From the resistivity we confirmed that  the clean model 
does not show
a thermally driven MIT except for $n \lesssim 0.1$, as 
noted  earlier \cite{hde-mc}. 

Even weak structural disorder  dramatically affects the
electronic state at strong coupling, leading to metal-insulator
transitions  which
are not present in the clean limit. 
In the rest of the paper we  focus on  $n=0.3$, 
typical of the high density
metallic regime, and study the effect of varying disorder and 
EP coupling.
Since the $r_A$ controlled bandwidth variation 
in real materials is  relatively small we use only 
a $\pm 10 \%$ variation
in $\lambda/t$ about  $\lambda=2.0$ 
and study the  weak disorder regime, $\Delta=0-1.0$. 

Disorder creates density inhomogeneities in the FL which are amplified by
the EP coupling, leading to a rapid rise in residual resistivity, and
also opens up a window for the crucial F-FL to P-PL crossover. 
Fig.2.$(a)-(c)$ show 
$\rho(T)$ with varying $\lambda$ and $\Delta$ illustrating the
three broad transport regimes.
At $\lambda = 1.8$,
which is towards the lower end of our coupling range,
moderate $\Delta$ increases $\rho(0)$ but the overall $\rho(T)$ has
the same nature as in the simple DE model \cite{sk-pm-scr} and
neither EP coupling nor disorder seems to have any qualitative effect.
This is true of $\Delta \lesssim 0.6$ and, we suggest, is akin to what
one observes in La$_{0.7}$Sr$_{0.3}$MnO$_3$.
 At $\Delta=1.0$, where 
 $\rho(0) \gtrsim 100 \sim
\rho_{Mott}$  we see hints of a weak downturn in $\rho(T)$ at high $T$. 
Overall, $\rho(T)$ 
at $\lambda=1.8$ is mainly characterised by a F-FL to
P-FL crossover. The density of states (DOS), 
Fig.2.$(d)$, remains featureless at all $T$
and the lattice distortions, Fig.2.$(g)$, are  weak
with no significant variation across $T_c$.
This ``DE like'' character in $\rho(T)$ survives upto  $\lambda=2.2$
in the clean limit, see Fig.2.$(b)-(c)$.
However, for $\lambda \gtrsim 2.0$ 
addition of weak disorder $(i)$~sharply increases $\rho(0)$ and
creates a pseudogap in the DOS, $(ii)$~maintains a regime with $d\rho/dT >0$
for $T \lesssim T_c$, but $(iii)$~crosses over to an `insulating' PL
regime with $d\rho/dT < 0$ for $T \gtrsim T_c$. As Fig.2.$(h)-(i)$ indicate, 
these
changes are associated with large lattice distortions and their variation
across $T_c$.
For  $\Delta  \sim 1.0$ the ground state itself turns insulating
for $\lambda \gtrsim 2.0$.  
We provide a compact picture of the various MIT in Fig.3, discussed
later.

To gain insight on the F-FL to P-PL crossover  
we also  calculated the `effective carrier
number',  $n_{eff}({\bar \omega}, T)
= \int_0^{\bar \omega} \sigma(\omega, T) d\omega$, at $T=0$ and ${\bar \omega}
=0.5$, as well the density field
$n({\bf r}, T)$. 
The combination of $n_{eff}$, Fig.4.$(a)$, the
spatial patterns (not shown here) and the DOS, Fig.2.$(e)$,
suggest the following scenario at strong coupling and moderate
disorder:
$(i)$~there is polaronic localisation of 
a {\it fraction} of carriers (states at $\omega < \mu$)
at $T=0$, lowering $n_{eff}$, 
but the states at the chemical potential remain
extended, $(ii)$~the 
polaronic states are further localised 
with increasing $T$
due to DE spin disorder, 
while the delocalised electrons near $\mu$ experience
scattering from  
the spin fluctuations, leading to $d\rho/dT >0$, 
$(iii)$~by the time $T \sim T_c$ the net  disorder 
arising from $\epsilon_i$, the 
lattice distortions, and  spin disorder is so large
that the system 
is effectively at a localisation  transition
on the complex $\{ \epsilon_i, x_i, {\bf S}_i \}$ background: 
the states at $\mu$ also get localised,
$(iv)$~the $T \gtrsim T_c$ phase 
has a modest  `mobility gap' (at moderate $\lambda, \Delta$) 
so there is no simple activated behaviour. 
The pseudogap in the DOS, Fig.2.$(e)$,  {\it deepens} initially and
gradually fills up for $T \gtrsim T_c$. 

Let us consider the relevance of these results to 
manganites at
$ x \sim 0.3-0.4$, away from commensurate filling and 
charge ordering instabilities.
The mean hopping 
amplitude between Mn ions, via the oxygen orbitals,
is controlled by  $r_A$ 
through the Mn-O-Mn bond angle. The  mismatch between
different A site cations is quantified 
through the variance, $\sigma_A$, of the ionic size, and acts as 
\begin{center}
\begin{figure}
\epsfxsize=7.8cm \epsfysize=9.9cm \epsfbox{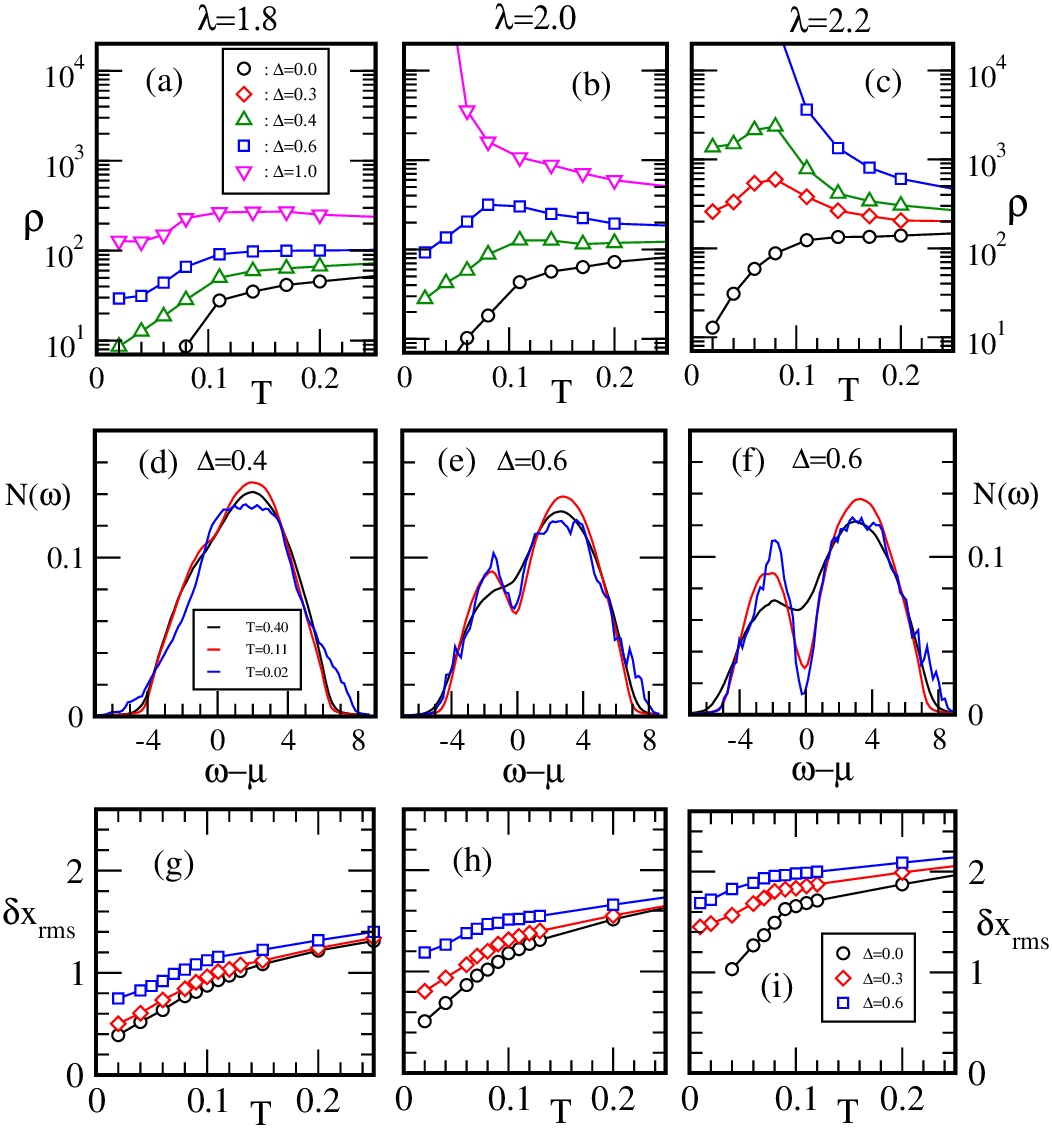}
\vspace{.1cm}
\caption{Colour online:
Resistivity, $\rho(T)$, DOS, and
lattice distortion  at $n=0.3$.
$(a)-(c).$  $\rho(T)$, with 
varying $\lambda$ and $\Delta$. 
$(d)-(f).$ DOS, $N(\omega,T)$ for the indicated parameters.
$(g)-(i).$~RMS lattice distortion, $\delta x_{rms}=
\sqrt{\langle (x_i - {\bar x})^2 \rangle}$,
where ${\bar x}$ is the system averaged distortion.}
\end{figure}
\end{center}
\begin{center}
\begin{figure}
\epsfxsize=7.5cm \epsfysize=7.5cm \epsfbox{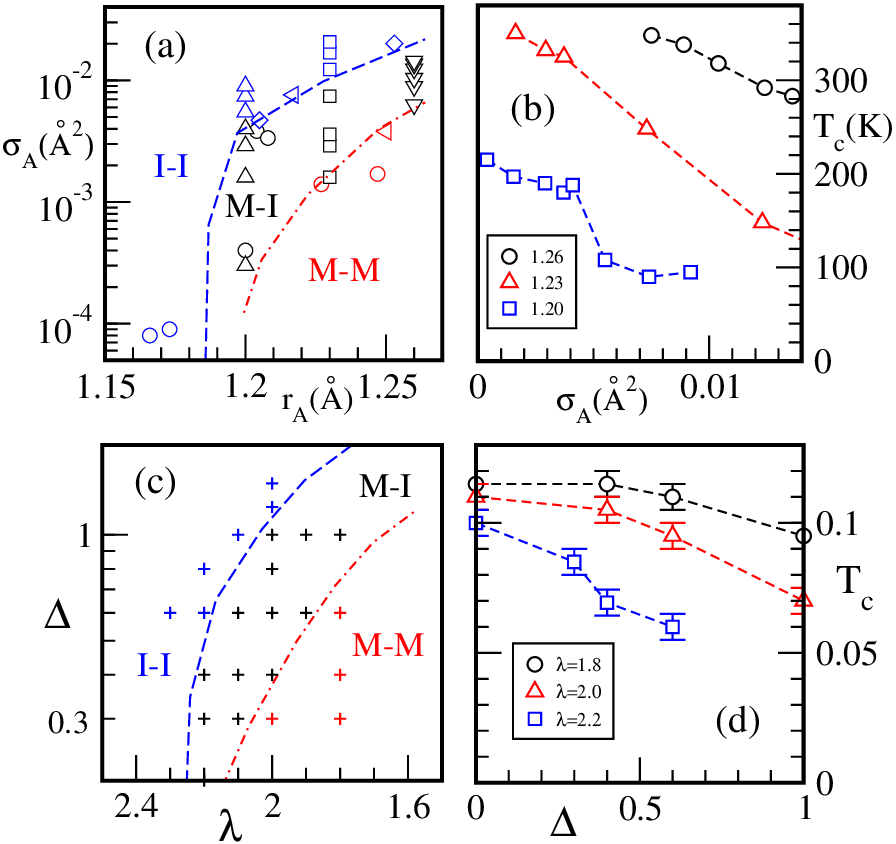}
\vspace{.1cm}
\caption{Colour online:
Transport regimes and ferromagnetic $T_c$: comparing
data on the $x=0.3$ manganites with our $n=0.3$ results.
$(a).$~Transport `phase diagram'
indicating the nature of low $T$ to high $T$ crossover with changing 
$r_A$ and $\sigma_A$, data from Ref[6-9] 
$(b).$~$T_c(\sigma_A, r_A)$ in the manganites, from Ref[6-7], 
$(c)$~transport crossovers from our calculation, and 
$(d).$~our results on $T_c(\Delta, \lambda)$. 
}
\end{figure}
\end{center}
disorder on
the electronic system. 
For a material of composition A$_{1-x}$A$'_x$MnO$_3$, say,
$r_A$ and $\sigma_A$  depends on the radius of the A, A' ions
as well as $x$.  
To disentangle the effect of varying
carrier density $(x)$ from that of varying `coupling constants', there
have been systematic studies of the manganites at fixed $x$, with
controlled variation in $r_A$, related to our $\lambda/t$ ratio, and 
$\sigma_A$, related to our~$\Delta/t$. 

Focusing initially on  $x=0.3$, the data from a wide variety of 
sources  can be organised in terms of the transport response,
$\rho(T)$, and magnetic $T_c$. We choose a 
simple characterisation where we construct the MIT 
phase diagram in terms of the low $T$ to high $T$ crossover 
observed in $\rho(T)$. For example,
La$_{0.7}$Sr$_{0.3}$MnO$_3$
(LaSr), is a metal at both low $T$ and $T > T_c$ so it shows M-M
crossover. In that spirit 
LaCa shows M-I  crossover, 
while PrCa is I-I.
These simple `binary' cationic systems are well studied, have 
relatively small $\sigma_A$,
and the
transport response can be argued to arise mainly from variations
in $r_A$. However, more complex cationic combinations
have been used, notably by Attfield and coworkers 
\cite{attf-lide,attf-sum}, 
to {\it vary $\sigma_A$ at fixed $r_A$ and $x$}. 
Our organisation of the
experimental data, Fig.3.$(a)$
 uses three such families, at $r_A = 1.26,~1.23$
and $1.20 \AA$, as well as data from  other sources \cite{liu,coey}
to highlight the combined impact of $r_A$ and 
$\sigma_A$
variation on transport character.
We have used the 9 fold coordinated values of ionic radius
from Shannon \cite{shannon} in computing 
$r_A$ and $\sigma_A$ for about 30 
compositions.

Four features are notable in the experimental 
MIT and $T_c$ data, Fig.3.$(a)-(b)$, at $x=0.3$: $(i)$~the `clean' limit  
allows only a narrow window,  $r_A \sim 1.18-1.20  \AA$,  for M-I 
\begin{center}
\begin{figure}
\epsfxsize=8.0cm \epsfysize=4.2cm \epsfbox{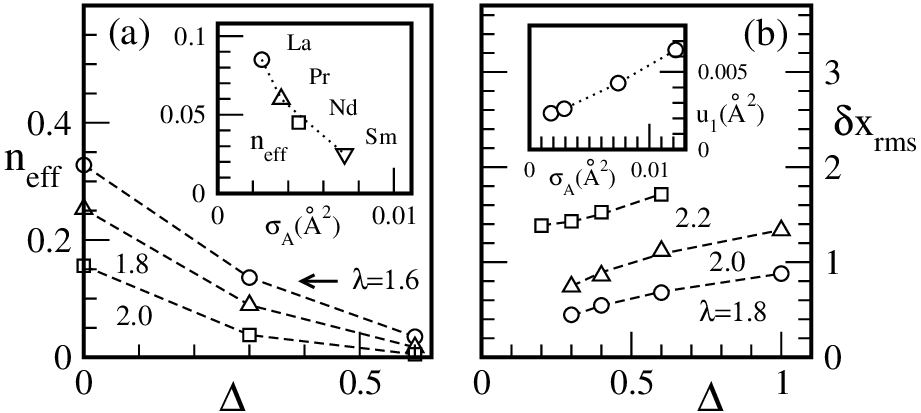}
\vspace{.1cm}
\caption{$(a)$~Computed optical spectral weight, 
$n_{eff}$, at cutoff $\omega = 0.5t$ and $T=0.01$ for varying $\Delta$ and
$\lambda$. Inset: data from Ref[10] on LaSr, {\it etc},
at $x=0.4$, $\omega=0.1$eV,
plotted versus $\sigma_A$. The $r_A$ values, LaSr downwards,
are 1.256 $\AA$, 1.232 $\AA$, 1.22 $\AA$ and 1.202 $\AA$.
$(b)$~Our results on disorder 
induced lattice distortion at $T=0.01$. Inset shows
the $\sigma_A$ induced increase in the mean square oxygen displacement
at low $T$ in a sequence of $x=0.3$ and $r_A = 1.23 \AA$ manganites,
from~Ref[6].}
\end{figure}
\end{center}
crossover, $(ii)$~the M-I region broadens with increasing
$\sigma_A$ at the cost of the M-M region (red/dot-dash line), 
disorder can {\it induce  a thermally driven MIT}, 
$(iii)$~at $T=0$
just increasing $\sigma_A$ 
can drive a metallic ground state insulating
(blue/dashed line), $(iv)$~the same $\sigma_A$ has much
stronger impact on transport and $T_c$ at small $r_A$ compared to
large $r_A$, for example $\sigma_A \sim 0.01 \AA^2$ has only
modest effect on $T_c$ at $r_A = 1.26 \AA$, but leads to a 
large reduction at $r_A \lesssim 1.23 \AA$. 
While we cannot make a direct numerical correspondence 
between $\sigma_A$ and $\Delta/t$, etc,  our 
results on transport crossovers, Fig.3.$(c)$, 
and $T_c$, Fig.3.$(d)$, show that {\it all
the trends $(i)-(iv)$ above} are correctly reproduced.
A  fit to the 
detailed $T$ dependence of
resistivity will of course require 
a more sophisticated phonon model.
Our $T_c^{max} \sim 0.12t
\sim 300$K (for $t \sim 0.25$eV) compares reasonably with the
maximum in  Fig.3.$(b)$. 

Although the similarity of our results to experimental data on
the manganites seems persuasive,  let us  
examine  how these results may be modified in a more realistic model. 
$(i)$~The manganites
involve {\it two $e_g$ orbitals} coupled to {\it Jahn-Teller} (JT)
phonons, with the possibility of
orbital order. Our preliminary calculations with
the 2 band JT model  in 2D \cite{2band-pap}
suggests that for $x \sim 0.3-0.4$ there is no non trivial orbital order
and the transport response is similar to what we observe here.
The one band model does not capture the correct physics as
$x \rightarrow 0$, or $x=0.5$, but appears  reasonable for $x \sim 0.3-0.4$.
$(ii)$~The oxygen atoms are corner shared among MnO$_6$ octahedra so
the {\it phonon variables are cooperative}. We have not explored this
aspect but believe the main effects would be 
lower critical EP coupling for localisation
\cite{hde-mc} and possibly  a sharper MIT across $T_c$.
$(iii)$~There are  {\it AF interactions} present in the 
manganites. We repeated the $d$-H-DE calculation in the presence of
AF coupling and discovered that it  
leads to  a sharper drop in $T_c$, shifts the MIT phase boundaries
somewhat, and leads to an AF-I phase at 
large $\lambda$ or $\Delta$. The  qualitative results of this paper,
however, remain unaffected.

Finally, if 
our argument about disorder induced polaron formation is
correct then the following features should be 
experimentally observed with increasing $\sigma_A$:
$(a)$~systematic suppression of low frequency
optical spectral weight, $n_{eff}$ at $\omega \lesssim 0.1$eV, at low
$T$, $(b)$~increase in the
incoherent lattice distortion (oxygen Debye-Waller factor),
progressively larger at smaller $r_A$,
and $(c)$~appearance of an increasingly  deeper 
pseudogap in the tunneling DOS at low $T$.
Fig.2 and Fig.4 quantify these effects.
Experiments in the recent past 
do claim disorder induced polaron formation in 
La$_{0.54}$Ba$_{0.46}$MnO$_3$
\cite{dis-pol-prl} but more direct confirmation  
for $x \sim 0.3$ is desirable.

To conclude, we have studied a  model with Holstein 
phonons, double exchange and quenched disorder, and   
illustrated the disorder enhancement of polaronic
tendency and its interplay with spin fluctuations.
Our results compare well with the disorder dependence of
transport and $T_c$ observed in
the `optimally doped' manganites. We make three testable 
predictions regarding the behaviour of optical spectral weight,
oxygen Debye-Waller factor and tunneling density of states 
for controlled variations of $\sigma_A$ in these materials.

We thank HRI for use of the Beowulf cluster  and  
Saurabh Madaan for analysis of the experimental data.

* Present address: Institute for Physics, 
Theoretical Physics III, Electronic
Correlations and Magnetism,
University of Augsburg, 86135 Augsburg, Germany.

{}

\end{document}